\documentclass[aps,showpacs,twocolumn,slashbox]{revtex4}
\usepackage{stmaryrd}
\usepackage{amsfonts}
\usepackage{mathrsfs}
\usepackage{amsmath}
\usepackage{amscd}
\usepackage{graphicx}
\usepackage{leftidx}
\begin{document}
\title{High-dimensional quantum state transfer through a quantum spin chain}
\author{Wei Qin$^{1,2}$}
\author{Chuan Wang$^3$}
\author{Gui Lu Long$^{1,2,}$}
\thanks{gllong@mail.tsinghua.edu.cn}
\address{$^1$ State Key Laboratory of Low-Dimensional Quantum Physics and Department of Physics, Tsinghua University, Beijing 100084,
China\\
 $^2$ Tsinghua National Laboratory for Information Science and Technology, Tsinghua
University, Beijing 100084, China\\
$^3$School of Science, Beijing University of Posts and Telecommunications, Beijing 100876,
China}
\begin{abstract}
In this paper, we investigate a high-dimensional quantum state transfer protocol. An arbitrary unknown high-dimensional state can be transferred with high fidelity between two remote registers through a XX coupling spin chain of arbitrary length. The evolution of the state transfer is determined by the natural dynamics of the chain without external modulation and coupling strength engineering. As a consequence, entanglement distribution with high efficiency can be achieved. Also the strong field and high spin quantum number can counteract partly the effect of finite temperature to ensure high fidelity of the protocol when the quantum data bus is in the thermal equilibrium state under an external magnetic field.
\end{abstract}
\maketitle

\section{introduction}

Quantum state transfer between two distant parties is an important task for quantum information processing (QIP).
The high fidelity state transfer relies on physical systems which can serve as quantum data buses to connect remote parties. There are many experimental realization of the data buses, such as phonons in ion traps \cite{ion1,ion2}, electrons in semiconductors \cite{semiconductor},
flux qubits in superconductors \cite{superconductor1,superconductor2,superconductor3}
and photons in optics \cite{photon1,photon2,photon3,photon4}. Recently, the solid-state spin chain system is becoming one of the most promising candidates for QIP due to its long decoherence time and ability to manipulate, transfer \cite{chain1,chain2,chain3,chain4,chain5,chain6,chain7}. The dynamics of such spin chain is determined by the evolution under a suitable Hamiltonian, e.g., the Heisenberg or XY Hamiltonian.

The first quantum state transfer (QST) protocol was proposed by Bose in which the spin state can be efficiently transferred through a spin chain via natural evolution \cite{Bose}. In the proposed protocol, two qubits are located at the two ends of the spin chain and the state of the encoded qubit at one end will be transferred to the other end after a specific mount of time without any operation on the chain. In the past decades, there are many QIP protocols based on the spin chain systems, especially the perfect quantum state transfer (PQST) \cite{perfect1,perfect2,perfect3,perfectl,perfect4,perfect5,perfect6,perfect7,perfect8,perfect9}. Later, Christandl et al. \cite{CH1,CH2} generalized the PQST of the spin chain to the spin network with arbitrary length based on the Cartesian product method of graph theory. Yao et al. \cite{N1,N2} proposed a high fidelity QST through an infinite temperature (unpolarized) quantum data bus.

Entanglement and quantum parallelism in quantum computation provide us the potential of merits superior to our conventional classical methods \cite{NC}. Moreover, it is difficult to raise the number of qubits coupled experimentally \cite{ST} and the high-dimensional systems can be coupled to a given dimensionality of the Hilbert space more efficiently using fewer systems than the two-dimensional systems. Hence, the high-dimensional systems can be employed as qudits to encode quantum information instead of qubits. The extensions of various protocols of quantum computation and communication
from two-dimensional systems to high-dimensional systems have been proposed, such as quantum
cloning \cite{clon1,clon2}, quantum cryptography \cite{cg}, quantum teleportation \cite{tele1,tele2},
quantum key distribution \cite{QKDd} and even implemented in experiment \cite{dexper}.
Additionally, the application of high-dimensional systems will enhance and deepen our understanding
of quantum computation and communication. The high-dimensional PQST through the spin chain
has been shown to be possible when the distances between the two end qudits are $1,2,4$ \cite{d}
and the three-dimensional case was realized as the distance is $3$ under
the bilinear-biquadratic Hamiltonian \cite{d3}.

In this paper, we propose an efficient protocol to achieve an arbitrary unknown high-dimensional QST through a XX coupling spin chain. The high $S$ spins are used to act as a quantum data bus and quantum registers. The low-lying level states of registers are used to encode quantum information as qudits. The data bus is initialized to a ferromagnetic order with the spins aligning in a parallel way. The states of the spin chain are mapped onto a set of bosons after Holstein-Primakoff transformation. Under the limit that the dimension of the sent state is much smaller than spin quantum number, the spin-wave interaction can be neglected to express the Hamiltonian of the chain in terms of free bosons and diagonalize the Hamiltonian of the data bus after an orthogonal transformation. Ensure that the register-bus coupling strength is much weaker than that between the data bus spins, the two registers are resonantly coupled to one of the collective eigenmodes of the data bus and other off-resonant coupling can be ignored. Then a swap gate between the two registers can be achieved at the optimal time. The proposed protocol requires neither external modulation of the Hamiltonian evolution nor spin chain coupling engineering.
Consequently, entanglement distribution with high efficiency can be achieved by using this scheme. We numerically simulate the average fidelity of the data bus in the thermal equilibrium state under an external magnetic field and find that the average fidelity decreases with temperature and increases with the field and spin quantum number.

The paper is organized as follows. In Section II, we give the XX model Hamiltonian and treat it.
In Section III, we show the high-dimensional
QST with high fidelity and entanglement distribution with high efficiency.
In Section IV, the case where the quantum data bus is in the thermal equilibrium state is investigated.
And the last section is our summary.

\section{model and calculation}

\begin{figure}[!ht]
\begin{center}
\includegraphics[width=8.5cm,angle=0]{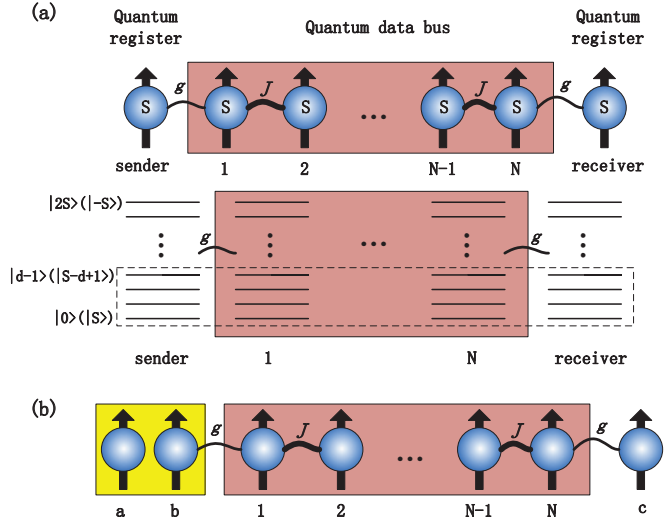}
\caption{(Color online)(a) Two distant quantum registers are intermediated by a quantum data bus with
the coupling strength $g$, which is much weaker than that between the data bus spins $J$.
By using the low-lying level states of the sender, ranging from the $|0\rangle$ state to the $|d-1\rangle$ state,
to encode as a qudit, we achieve a high-dimensional QST with high fidelity.
(b) Entanglement distribution through a spin chain. The sites $b$
and $c$ are coupled to the quantum data bus, while the site $a$ not. $a$ and $b$
are prepared in a maximally entangled state.
Under the natural evolution $a$ and $c$
gain entanglement from the maximally entangled
state with high efficiency at the optimal time.}\label{f1}
\end{center}
\end{figure}

A spin chain consists of $N$ sites, where each site comprises a single $S$ spin in the state $| m \rangle$ ($m=S,S-1,...,-S+1,-S$). The spin chain is used to act as a quantum data bus. The data bus is placed between two additional spins, which are considered as two quantum registers, denoted as the sender $s$ and the receiver $r$. The Hamiltonian of the system consisting of spins coupled to their nearest neighbors on a finite lattice of site $N+2$ as shown in figure \ref{f1}(a) includes two terms (We have set $\hbar=1$ in the present paper):
\begin{equation}\label{eq1}
H=H_{XX}+H_{M},
\end{equation}
where $H_{XX}=H_{B}+H_{I}$. $H_{B}$ is the XX coupling of the data bus as
\begin{equation}\label{eq2}
H_{B}=-J\sum_{(i,j)}(S^{+}_{i}S^{-}_{j}+S^{-}_{i}S^{+}_{j}),
\end{equation}
where $J>0$ is the coupling strength between the data bus spins, $(i,j)$ means that only
the nearest neighbor (NN) coupling is considered.
$S^{\pm}_{i}=S^{x}_{i} \pm iS^{y}_{i}$ and $S^{\nu}_{i}$ ($\nu=x,y,z$)
is the $\nu$ component of the spin operator $\textbf{S}_{i}$ at the site $i$.
The interaction Hamiltonian between the two registers and data
bus has the form
\begin{equation}\label{eq3}
H_{I}=-g(S^{+}_{s}S^{-}_{1}+S^{+}_{r}S^{-}_{N}+H.C.),
\end{equation}
where $g>0$ is the register-bus coupling strength and $H.C.$ denotes the complex conjugate.
$H_{M}$ is the Zeeman term under an external magnetic field $h$ given by
\begin{equation}\label{HM}
H_{M}=-h(S^{z}_{s}+S^{z}_{r}+\sum_{i=1}^{N}S^{z}_{i}).
\end{equation}
After the Holstein-Primakoff (HP) transformation \cite{HP}
\begin{eqnarray}\label{eq4}
&&S^{+}_{i}=\sqrt{2S-a^{\dag}_{i}a_{i}}a_{i},\nonumber\\
&&S^{-}_{i}=a^{\dag}_{i}\sqrt{2S-a^{\dag}_{i}a_{i}},\nonumber\\
&&S^{z}_{i}=S-a^{\dag}_{i}a_{i},
\end{eqnarray}
with $[a_{i},a_{j}^{\dag}]=\delta_{ij}$ and $[a^{\dag}_{i},a^{\dag}_{j}]=[a_{i},a_{j}]=0$,
the two Hamiltonians $H_{XX}$ and $H_{M}$ are expressed in terms of the bosonic operators.

Initially, the data bus and receiver has a simple ferromagnetic order with the spins aligning in a parallel way
denoted by $|0\rangle_{bus}^{\otimes N}|0\rangle_{r}$ with $|0\rangle_{bus}^{\otimes N}=|0\rangle_{1}...|0\rangle_{N}$.
The low-lying level states of the sender, ranging from the $|0\rangle$ state to the $|d-1\rangle$ state,
are employed to encode information as a qudit. The total boson number,
\begin{equation}\label{eq22}
N_{tot}=N_{s}+N_{r}+\sum^{N}_{i=1}N_{i},
\end{equation}
is conserved. The dimension of the Hilbert space $\mathcal{H}$ associated with
the spin-$S$ chain of length $N+2$ is $(2S+1)^{N+2}$. The dimension of the state that should be sent is $d$
and the conservation
law ensures that the $d$-dimensional state transfer dynamics is determined by the
evolution in the $d^{N+2}$-dimensional subspace $\mathcal{S}$ spanned by the basis vectors
$|n_{s}n_{1}...n_{N}n_{r}\rangle$ ($n_{i}=0,...,d-1$).
We assume that the dimension $d$ of the sent state
is much smaller than $2S$, as $d \ll 2S$, which gives $\langle a^{\dag}_{i}a_{i} \rangle \ll 2S$.
The HP transformation is simplified to \cite{low}
\begin{equation}\label{SHP}
S^{+}_{i}=a_{i}\sqrt{2S},S^{-}_{i}=a_{i}^{\dag}\sqrt{2S}.
\end{equation}

Substituting Eq. (\ref{SHP}) into Eqs. (\ref{eq2}), (\ref{eq3}) and (\ref{HM}),
we have
\begin{equation}\label{eq8}
H_{B}=-2SJ\sum_{(i,j)}(a^{\dag}_{i}a_{j}+a_{i}a^{\dag}_{j}),
\end{equation}
\begin{eqnarray}\label{eq9}
&&H_{I}=-2Sg(a^{\dag}_{s}a_{1}+a^{\dag}_{r}a_{N}+H.C.)
\end{eqnarray}
and
\begin{equation}
H_{M}=-(N+2)hS+h(a^{\dag}_{s}a_{s}+a^{\dag}_{r}a_{r}+\sum^{N}_{i=1}a^{\dag}_{i}a_{i}).
\end{equation}
Here the spin-wave interaction of the system is neglected and the three Hamiltonians are reexpressed
in terms of free bosons.

As $g/J \ll 1$, $H_{B}$ works as a collective Hamiltonian and $H_{I}$ works as a perturbation one, the diagonalization of the Hamiltonian $H_{B}$ of Eq. (\ref{eq8}) occurs after the following orthogonal transformation\cite{N1,N2,XXd}
\begin{eqnarray}\label{eq10}
a^{\dag}_{i}=\frac{1}{A}\sum_{k=1}^{N}\sin \frac{ik\pi}{N+1} b_{k}^{\dag},
\end{eqnarray}
with $k=1,...,N$ and $A=\sqrt{\frac{N+1}{2}}$, the Hamiltonian $H_B$ becomes
\begin{equation}\label{eq11}
H_{B}=\sum^{N}_{k=1}\varepsilon_{k}b^{\dag}_{k}b_{k},
\end{equation}
where $\varepsilon_{k}=-4SJ\cos(\frac{k\pi}{N+1})$.
Combining Eqs. (\ref{eq9}) and (\ref{eq10}), it gives that
\begin{eqnarray}\label{eq12}
H_{I}=\sum^{N}_{k=1}t_{k}(a_{s}b^{\dag}_{k}+(-1)^{k-1}a_{r}b^{\dag}_{k}+H.C.)
\end{eqnarray}
with $t_{k}=-\frac{2Sg}{A}\sin\frac{k\pi}{N+1}$.

\section{Quantum state transfer and entanglement distribution}
\subsection{Quantum state transfer}

In this section, the high-dimensional QST with high fidelity through a spin chain in the absence of an external magnetic field is proposed. If the spin number in the data bus is odd, there is a zero energy bosonic eigenmode corresponding to $k=\kappa\equiv (N+1)/2$. By maintaining $g/J\ll 1$ to ensure $t_{\kappa}\ll |\varepsilon_{\kappa}-\varepsilon_{\kappa\pm1}|$, the off-resonant coupling to other bosonic eigenmodes can be neglected and the two end registers are resonantly coupled to the $\kappa th$ eigenmode \cite{N1,N2} with $t_{\kappa}=-2Sg/A$.
The state transfer dynamics is drived by the evolution under the effective Hamiltonian
\begin{equation}\label{eq14}
H_{eff}=t_{\kappa}(a^{\dag}_{s}b_{\kappa}+(-1)^{\kappa-1}a^{\dag}_{r}b_{\kappa}+H.C.).
\end{equation}

In Heisenberg picture, the evolution of the operator is determined by $O(\tau)=U^{\dag}(\tau)O(0)U(\tau)$, where $U(\tau)=e^{-iH\tau}$ is the evolution operator. The operators $A$ and $B$ obey the rule that
$e^{-\alpha A}Be^{\alpha A}=B-\alpha[A,B]+\frac{\alpha^{2}}{2!}[A,[A,B]]+...$. By applying the relations
we get
\begin{eqnarray}\label{eq15}
&&U_{eff}^{\dag}a_{s}^{\dag}U_{eff}=a_{s}^{\dag}+\frac{1}{2}(a_{s}^{\dag}+(-1)^{\kappa-1}a_{r}^{\dag})\nonumber\\
&&[-1+\cos(\sqrt{2}t_{\kappa}\tau)]
+ib_{\kappa}^{\dag}\frac{\sin(\sqrt{2}t_{\kappa}\tau)}{\sqrt{2}}.
\end{eqnarray}
At the optimal time $\tau=\tau_{0}\equiv\pi/\sqrt{2}t_{\kappa}$, we have
\begin{equation}\label{eq16}
U_{eff}^{\dag}a_{s}^{\dag}U_{eff}|_{\tau_{0}}=(-1)^{\kappa}a_{r}^{\dag}.
\end{equation}
Similarity,
\begin{equation}\label{eq17}
U_{eff}^{\dag}a_{r}^{\dag}U_{eff}|_{\tau_{0}}=(-1)^{\kappa}a_{s}^{\dag}.
\end{equation}
Eqs. (\ref{eq16}) and (\ref{eq17}) show that the creation operator referring to the
sender (receiver) at $\tau=0$ becomes that referring to the receiver (sender) at the optimal time $\tau_{0}$.
Actually, it is a swap gate between the two registers with an additional phase $(-1)^{\kappa}$.
The additional phase is independent of the sent state and
only determined by the length of the data bus. Thus the receiver will obtain an
arbitrary unknown state that is sent at initial time after a phase gate operation
given by
\begin{equation}\label{eq22a}
P_{N+1}^{d}=\left(
\begin{array}{ccccc}
    1 & 0 & 0 & \cdots & 0  \\
    0 & (-1)^{\kappa} & 0 & \cdots & 0 \\
    0 & 0 & (-1)^{2\kappa} & \cdots & 0 \\
    \vdots & \vdots & \vdots & \ddots & \vdots \\
    0 & 0 & 0 &\cdots & (-1)^{(d-1)\kappa} \\
\end{array}
\right)_{N+1}^{d}.
\end{equation}

For simplicity, we take the $4$-dimensional state transfer as an example.
The sent initial state is $|\varphi\rangle_{s}
=\sum_{\mu=0}^{3}\alpha_{\mu}|\mu\rangle
=\sum_{\mu=0}^{3}\frac{\alpha_{\mu}}{\sqrt{\mu!}}(a_{s}^{\dag})^{\mu}|0\rangle$
and $\sum_{\mu=0}^{3}|\alpha_{\mu}|^{2}=1$.
The initial state of the whole spin chain, including the sender, receiver and data bus,
shall be
\begin{equation}\label{eq18}
|\psi(0)\rangle=[\sum_{\mu=0}^{3}\frac{\alpha_{\mu}}{\sqrt{\mu!}}(a_{s}^{\dag})^{\mu}|0\rangle]_{s}
|0\rangle_{bus}^{\otimes N}|0\rangle_{r}.
\end{equation}
At the optimal time $\tau_{0}$, the creation operator referring to the
sender becomes that referring to the receiver, we obtain the final state
\begin{eqnarray}\label{eq19}
&&|\psi(\tau_{0})\rangle\nonumber\\
&&=|0\rangle_{s}|0\rangle_{bus}^{\otimes N}[\sum_{\mu=0}^{3}\frac{\alpha_{\mu}}{\sqrt{\mu!}}
(-1)^{\mu \kappa}(a_{r}^{\dag})^{\mu}|0\rangle]_{r}\nonumber\\
&&=|0\rangle_{s}|0\rangle_{bus}^{\otimes N}[\sum_{\mu=0}^{3}(-1)^{\mu \kappa}\alpha_{\mu}|\mu\rangle]_{r}.
\end{eqnarray}
After a phase gate operation
\begin{equation}\label{eq20}
P_{N+1}^{4}=\left(
\begin{array}{cccc}
    1 & 0 & 0 & 0 \\
    0 & (-1)^{\kappa} & 0 & 0 \\
    0 & 0 & (-1)^{2\kappa} & 0 \\
    0 & 0 & 0 & (-1)^{3\kappa} \\
\end{array}
\right)_{N+1}^{4},
\end{equation}
we have $P_{N+1}^{4}|\psi(\tau)\rangle=|0\rangle_{s}|0\rangle_{bus}|\varphi\rangle_{r}$.

In two-dimensional Hilbert space, the set of the pure states forms a complex projective space
$\mathbb{C}P^{1}$, one can use the two-dimensional Bloch sphere to measure these pure states, since
in this case $U(2)/U(1)\sim SO(3)/SO(2)\sim S_{2}$. The set of the pure states of a $d$-dimensional
Hilbert space constructs a complex projective space $\mathbb{C}P^{d-1}$, where the natural uniform measure
exists. To measure a random pure state on this $2(d-1)$-dimensional manifold, a vector of
a random unitary matrix distributed over the invariant (Harr) measure on $U(d)$ is taken. A $d$-dimensional
normalized pure state $|\phi\rangle$ can be measured by the Hurwitz
parametrization with $d-1$ polar angles $\chi_{p}$ and $d-1$ azimuthal angles $\theta_{p}$ as \cite{HBS}
\begin{eqnarray}\label{hur}
&&|\phi\rangle \nonumber\\
&&=(\cos\theta_{d-1},\sin_{d-1}\cos\theta_{d-2}e^{i\chi_{d-1}},
\sin\theta_{d-1}\sin\theta_{d-2}\nonumber\\
&&\cos\theta_{d-3}e^{i\chi_{d-2}},...,\prod_{i=1}^{d-1}\sin\theta_{i}e^{i\chi_{1}}),
\end{eqnarray}
where $0\leq \theta_{p} \leq \pi/2$ and $0\leq \chi_{p} < 2\pi$ with $p=1,2,...,d-1$.
In analogy to the volume element on the two-dimensional Bloch sphere, that on
the generalized Bloch sphere of Eq. (\ref{hur}) in the complex space $\mathbb{C}P^{d-1}$ can be
written as
\begin{equation}
dV=\prod^{d-1}_{p=1}\cos\theta_{p}(\sin\theta_{p})^{2p-1}d\theta_{p}d\chi_{p},
\end{equation}
the total volume of the $2(d-1)$-dimensional manifold of the pure states is $V_{d}=\pi^{d-1}/(d-1)$.
The average of an observable $O$ over the whole manifold
of the pure states is $\langle O \rangle=\int_{\mathbb{C}P^{d-1}} O dV/V_{d}$.

The sent state is $|\varphi\rangle_{s}
=\sum_{\mu=0}^{d-1}\alpha_{\mu}|\mu\rangle$ and the initial state of the entire spin
chain is
\begin{eqnarray}\label{IST}
|\psi(0)\rangle=\sum_{\mu=1}^{d-1}\alpha_{\mu}|\mu\rangle_{s}|0\rangle^{\otimes N}_{bus}|0\rangle_{r}.
\end{eqnarray}
The state at time $\tau$ shall be
\begin{eqnarray}
|\psi(\tau)\rangle=\sum_{(n_{s}n_{1}...n_{N}n_{r})}c_{(n_{s}n_{1}...n_{N}n_{r},\tau)}
|n_{s}n_{1}...n_{N}n_{r}\rangle
\end{eqnarray}
and
\begin{equation}
c_{(n_{s}n_{1}...n_{N}n_{r},\tau)}=\sum_{\mu=1}^{d-1}\alpha_{\mu}
f_{(n_{s}n_{1}...n_{N}n_{r})}^{\mu},
\end{equation}
where $f_{(n_{s}n_{1}...n_{N}n_{r})}^{\mu}=\langle n_{s}n_{1}...n_{N}n_{r}
|e^{-iH\tau}|\mu\rangle_{s}|0\rangle^{\otimes N}_{bus}|0\rangle_{r}$. The state of the receiver at $\tau$ is denoted
by $\rho_{r}(\tau)$ since it is generally a mixed state. It can be obtained by tracing out
the other sites
\begin{eqnarray}
&&\rho_{r}(\tau)\nonumber\\
&&=tr_{\hat{r}}(|\psi(\tau)\rangle\langle\psi(\tau)|)\nonumber\\
&&=\sum_{n_{r},n_{r}'=0}^{d-1}\beta_{(n_{r}n_{r}',\tau)}|n_{r}\rangle\langle n_{r}'|
\end{eqnarray}
with
\begin{equation}
\beta_{(n_{r}n_{r}',\tau)}
=\sum_{(n_{s}^{''}n_{1}''...n_{N}'')}c_{(n_{s}''n_{1}''...n_{N}''n_{r},\tau)}
c^{*}_{(n_{s}''n_{1}'',...,n_{N}''n_{r}',\tau)}.
\end{equation}

The fidelity between the sent initial state $|\varphi\rangle_{s}$
and the received final state $\rho_{r}(\tau)$ is defined by
$F(\tau)=\leftidx{_s}\langle\varphi|\rho_{r}(\tau)|\varphi\rangle_{s}$, which turns out to be
\begin{equation}
F(\tau)=\sum_{\mu,\mu'=0}^{d-1}\alpha_{\mu'}\alpha_{\mu}^{*}\beta_{(\mu\mu',\tau)}.
\end{equation}
The average of the fidelity over the complex projective space $\mathbb{C}P^{d-1}$ is
\begin{equation}
\langle F(\tau) \rangle=\frac{1}{V_{d}}\int_{\mathbb{C}P^{d-1}} F(\tau) dV.
\end{equation}

\begin{figure}[!ht]
\begin{center}
\includegraphics[width=7cm,angle=0]{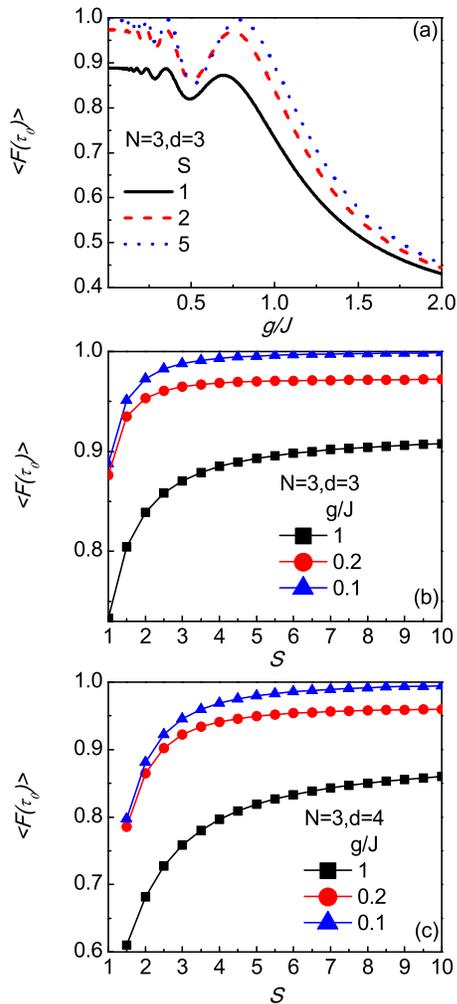}
\caption{(Color online)The average fidelity at the optimal time $\tau_{0}$.
(a) The average fidelity as a function of $g/J$
at $N=3$ and $d=3$.
The average fidelity as a function of spin quantum number at
(b) $N=3$, $d=3$ and (c) $N=3$, $d=4$.}\label{f2}
\end{center}
\end{figure}

Figure \ref{f2} shows the numerical results of the average fidelity
at the optimal time $\tau_{0}$.
The average fidelity as a function of $g/J$ is depicted in figure \ref{f2}(a).
$\langle F(\tau_{0})\rangle$ decreases with either $g/J$ increasing
or spin quantum number $S$ decreasing
when $g/J\geq 1$. As $g/J\rightarrow 0$, it reaches its maximum value, which increases
with spin quantum number. Noted that $\langle F(\tau_{0})\rangle$ could reach high values
at some $g/J$ values which are not much smaller than unity for high spin quantum number.
The two cases where $N=3$, $d=3$ and $N=3$, $d=4$ for three $g/J$ values
are depicted in figures \ref{f2}(b) and \ref{f2}(c), respectively.
It is seen that $\langle F(\tau_{0})\rangle$ increases with spin quantum number $S$.
As $g/J\ll1$ and
$d\ll2S$, $\langle F(\tau_{0})\rangle$ shall tend to $1$, for example, $\langle F(\tau_{0})\rangle=0.9987$
in the case $N=3$, $d=3$, $g/J=0.1$, $S=10$ and
$\langle F(\tau_{0})\rangle=0.9946$ in the case $N=3$, $d=4$, $g/J=0.1$, $S=10$.

\subsection{Entanglement distribution}

Entanglement has been considered as an important resource in QIP \cite{NC}. And entanglement distribution
among involved parties is also required in quantum teleportation \cite{QT}, quantum dense coding \cite{QDC} and quantum key distribution \cite{QKD}.
The proposed model can also be generalized to complete entanglement distribution between two remote parties.

We assume that the sites $b$ and $c$ are coupled to the quantum data bus but site a not as seen in figure 1(b).
At initial time, $a$ and $b$ are prepared in a maximally entangled state
\begin{equation}\label{LNI}
|\varphi\rangle_{ab}=\frac{1}{\sqrt{d}}\sum_{\mu=0}^{d-1}
|\mu\rangle_{a} |\mu\rangle_{b}.
\end{equation}
The purpose is to transfer
this entanglement by the natural dynamics of the data bus to achieve a maximally entangled state between $a$ and $c$.

The state of the composite system can be described as
\begin{equation}\label{LNWI}
|\psi(0)\rangle=\frac{1}{\sqrt{d}}\sum^{d-1}_{\mu=0}|\mu\rangle_{a}[\frac{(a^{\dagger}_{b})^{\mu}}
{\sqrt{\mu!}}|0\rangle_{b}|0\rangle^{\otimes N}_{bus}|0\rangle_{c}].
\end{equation}
After evolution, the finial state at the time $\tau_{0}$ is
\begin{equation}
|\psi(\tau_{0})\rangle=|0\rangle_{b}|0\rangle^{\otimes N}_{bus}
[\frac{1}{\sqrt{d}}\sum_{\mu=0}^{d-1}(-1)^{\kappa \mu}|\mu\rangle_{a}|\mu\rangle_{c}].
\end{equation}
The maximally entangled state between $a$ and $c$ has been achieved. The logarithmic negativity
\cite{LN1,LN2} is used to measure the entanglement of the state $\rho_{12}$, which is given by
\begin{equation}\label{LN}
LE_{12}\equiv log_{2}\parallel\rho^{T_{1}}_{12}\parallel.
\end{equation}
Here $T_{1}$ denotes the partial transpose of the density matrix $\rho_{12}$ with respective
to the subsystem $1$ as
\begin{equation}
\langle i_{1},j_{2}|\rho^{T_{1}}_{12}|k_{1},l_{2}\rangle
=\langle k_{1},j_{2}|\rho_{12}|i_{1},l_{2}\rangle
\end{equation}
The trace norm of operator $O$ is $\parallel O \parallel\equiv tr\sqrt{O^{\dag}O}=1+2|\sum_{i}\lambda_{i}|$,
$\lambda_{i}$ is the negativity eigenvalue of $\rho^{T_{1}}_{12}$.
For a pure maximally entangled state of Eq. (\ref{LNI}), the logarithmic negativity
is yielded that $LE_{ab}=log_{2}d$ from Eq. (\ref{LN}).
The efficiency of entanglement distribution \cite{d} is defined by
\begin{equation}
E=\frac{LE_{ac}}{LE_{ab}},
\end{equation}
which is used to measure the entanglement between $a$ and $c$ gained from
the maximally entangled state between $a$ and $b$ under the spin chain evolution.

Figure 3 demonstrates
the efficiency of entanglement transition through a data bus of length $N=3$ as a function of
spin quantum number at the optimal time $\tau_{0}$. It is obvious that the
efficiency increases with spin quantum number and approaches to unity when spin quantum number
is high enough. For example, $E(\tau_{0})=0.9984$ at $d=3$, $S=10$ and $E(\tau_{0})=0.9948$
at $d=4$, $S=10$.
\begin{figure}[!ht]
\begin{center}
\includegraphics[width=7cm,angle=0]{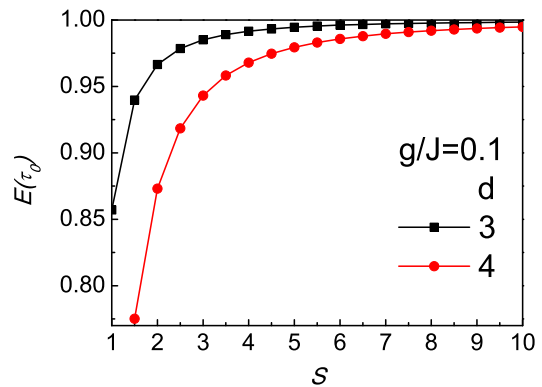}
\caption{(Color online)The efficiency of entanglement transition
through a data bus of length $N=3$ when $g/J=0.1$ at the optimal time $\tau_{0}$.}\label{f3}
\end{center}
\end{figure}

\section{Quantum data bus in the thermal equilibrium state}

In the following, we will investigate the effects of temperature on the quantum data
bus which is in the thermal equilibrium state in the presence of an external magnetic field $h$.

In the case where $h\neq 0$, the evolution operator can be described as
\begin{equation}\label{MEO}
e^{-iH\tau}=e^{-iH_{XX}\tau}e^{-iH_{M}\tau}
\end{equation}
since $[H_{XX},H_{M}]=0$. The finial state at the optimal time $\tau_{0}$ under the evolution operator of Eq. (\ref{MEO}) starting
from the initial state of Eq. (\ref{IST}) is
\begin{equation}
|\psi(\tau)\rangle=|0\rangle_{s}|0\rangle_{bus}^{\otimes N}
\sum^{d-1}_{\mu=0}(-1)^{\kappa \mu}e^{-ih\mu \tau_{0}}\alpha_{\mu}|\mu\rangle,
\end{equation}
where the overall phase induced by $-(N+2)hS$ has been neglected. An additional phase $e^{-ih\mu \tau_{0}}$ is introduced by the field. When the quantum
data bus is in the thermal equilibrium state, the density matrix satisfies the Boltzmann distribution as
\begin{eqnarray}
\rho_{B}=\frac{1}{Z}e^{-\frac{H_{B}}{T}}
=\frac{1}{Z}\sum_{i}e^{-\frac{E_{i}}{T}}|\phi_{i}\rangle\langle\phi_{i}|,
\end{eqnarray}
where $|\phi_{i}\rangle$ and $E_{i}$ are the eigenvectors and eigenvalues of the data bus, respectively.
$Z=tr(e^{-H/T})$ is the partial function and $T$ represents the temperature. The Boltzmann constant
$k_{B}$ has been set to be unity. As $h=0$, the state $|0\rangle^{\otimes N}_{bus}$
with zero energy is not the ground state of the data bus. As $h\neq 0$, a magnetic field is added
into Eq. (\ref{eq11}) and it becomes
\begin{equation}
H_{B}=\sum^{N}_{k=1}(\varepsilon_{k}+h)b^{\dag}_{k}b_{k}-hNS.
\end{equation}
The eigenvalue of the state $|0\rangle^{\otimes N}_{bus}$ is
$-hNS$. Hence, $|0\rangle^{\otimes N}_{bus}$ is the ground state as
$h \geq 4SJ \cos \frac{\pi}{N+1}$. At zero temperature the data bus
is in $|0\rangle^{\otimes N}_{bus}$ with no excited boson, and the data bus spins are
in a ferromagnetic order. With the increase of temperature, the number of the
excited bosons becomes larger.

\begin{figure}[!ht]
\begin{center}
\includegraphics[width=7cm,angle=0]{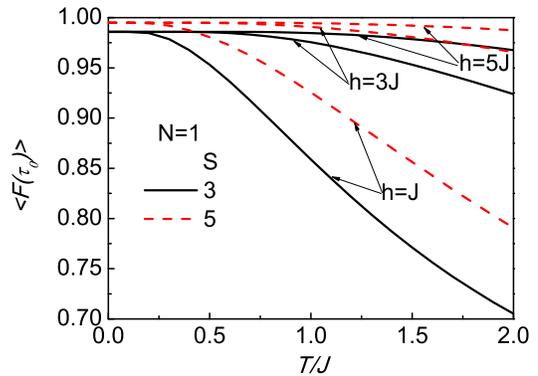}
\caption{(Color online)The average fidelity as a function of temperature
when $N=1$, $d=3$ in the cases $S=3$ and $5$ at the optimal time $\tau_{0}$.}\label{f4}
\end{center}
\end{figure}

The initial state of the whole spin chain is
\begin{equation}
\rho(0)=\sum^{d-1}_{\mu,\mu'=0}\alpha_{\mu}\alpha_{\mu'}^{*}|\mu\rangle_{s}
|0\rangle_{r}\rho_{B}\leftidx{_s}\langle \mu'| \leftidx{_r}\langle 0|,
\end{equation}
when the data bus is in the thermal equilibrium state.
The state of the reduced density matrix of the receiver at time $\tau$ is $\rho_{r}(\tau)=
tr_{\hat{r}}[e^{-iH\tau}\rho(0)e^{iH\tau}]$. Figure \ref{f4} depicts the average fidelity
as a function of temperature at the optimal time $\tau_{0}$ when the data bus is in the thermal equilibrium state
for several fields.

Figure \ref{f4} shows that the average fidelity decreases with temperature and increases with the field.
With the temperature increasing from zero, the number of the excited bosons increases in the data bus,
which could not ensure the condition $\langle a^{\dag}_{i}a_{i}\rangle \ll 2S$, to
lower the average fidelity. On the other hand, the field depresses the excitation of
the bosons to reduce the effect of temperature and raise the average fidelity. A feature of figure \ref{f4} is that
all red curves with $S=5$ vary not so steeply as the corresponding black ones with $S=3$,
which means that the high spin quantum number can counteract partly the effect of temperature.
Therefore, at finite temperature the strong field and high spin quantum number can
ensure high fidelity of our protocol.


\section{Summary}

In this paper, we have demonstrated that an arbitrary unknown high-dimensional state can be transferred between two parties at the optimal time with high fidelity by arbitrary length.
Consequently, the entanglement distribution between two remote parties can be realized with high efficiency.
This protocol does not require external modulation and coupling engineering.
Its a direct application is to communicate between two remote registers in a quantum computer using high-dimensional
systems, which can construct a bigger Hilbert space to process much more information than two-dimensional ones.
When the quantum data bus is in the thermal equilibrium state under an external magnetic field, the average fidelity decreases with temperature and increases with either the field or spin quantum number. Thus both the strong field and high spin quantum number can ensure high fidelity of our protocol. This helps the protocol to be able to work in the thermal environment.

\section{Acknowledgement}

This work was supported by the National Natural Science Foundation of China (Grant No. 61205117,11175094), the National Basic Research Program of China (2009CB929402, 2011CB9216002)£¬and Specialized Research Fund for the Doctoral
Program of Education Ministry.

\end{document}